\documentclass[prb,showpacs,twocolumn,preprintnumbers,amsmath,groupedaddress,amssymb,superscriptaddress,floatfix]{revtex4}
\usepackage[dvips]{graphicx}
\usepackage[latin1]{inputenc}
\usepackage{xcolor,bm}
\usepackage{nicefrac}
\usepackage{rotating}

\usepackage{stmaryrd}
\usepackage{color}
\usepackage{bm}      % RevTex4 bold math
\usepackage[normalem]{ulem}

\DeclareGraphicsExtensions{.eps,.png,.pdf}
\usepackage{dcolumn}           % For aligning table column on decimal point
\usepackage{nicefrac}          % For nice small fractions

\newcommand{\lafemnasof}{LaFe$_{1-x}$Mn$_{x}$AsO$_{0.89}$F$_{0.11}$ }

\newcommand{\mossbauer}{M\"{o}ssbauer }
\newcommand{\lnfemnasof}{LnFe$_{1-x}$Mn$_{x}$AsO$_{0.89}$F$_{0.11}$ }

\begin{document}
\title{Fast recovery of the pristine magnetic and structural phases in superconducting La\-Fe\-As\-O$_{0.89}$\-F$_{0.11}$ by Mn/Fe substitution}

\author{S.~Sanna}\email[E-mail: ]{s.sanna@unibo.it}\affiliation{Dipartimento di Fisica e Astronomia, Universit\`a di Bologna, via Berti-Pichat 6-2, I-40127 Bologna, Italy}\affiliation{CNR-SPIN and Universit\`a di Genova, via Dodecaneso 33, I-16146 Genova, Italy}
\author{P.~Carretta}\affiliation{Dipartimento di Fisica and Unit\`a CNISM di Pavia, I-27100 Pavia, Italy}
\author{M.~Moroni}\affiliation{Dipartimento di Fisica and Unit\`a CNISM di Pavia, I-27100 Pavia, Italy}
\author{G.~Prando}\affiliation{Dipartimento di Fisica and Unit\`a CNISM di Pavia, I-27100 Pavia, Italy}
\author{P.~Bonf\`a}\affiliation{CINECA, Casalecchio di Reno 6-3, 40033 Bologna, Italy}
\author{G.~Allodi}\affiliation{Dipartimento di Scienze Matematiche, Fisiche e Informatiche, Universit\'a di Parma, I-43124 Parma, Italy}
\author{R.~De~Renzi}\affiliation{Dipartimento di Scienze Matematiche, Fisiche e Informatiche, Universit\'a di Parma, I-43124 Parma, Italy}
\author{A.~Martinelli}\affiliation{CNR-SPIN and Universit\`a di Genova, via Dodecaneso 33, I-16146 Genova, Italy}
\begin{abstract}
We report on an experimental study of the effect of Mn impurities in the optimally doped La\-Fe\-As\-O$_{0.89}$\-F$_{0.11}$ compound. The results show that a very tiny amount of Mn, of the order of 0.1\%, is enough to destroy superconductivity and to recover at low temperatures both the magnetic ground state and the orthorhombic structure of the pristine LaFeAsO parent compound. The results are discussed within a model where electron correlations enhance the Ruderman-Kittel-Kasuya-Yosida interaction among impurities.
\end{abstract}

\maketitle

\section{Introduction}\label{sec:intro}
The study of the effect of impurities in superconductors is
relevant both in view of their technological applications \cite{Ma2012,Hosono2018,Prando2012}as well
as for the understanding of the microscopic mechanisms driving
superconductivity \cite{Sato2010a,Sato2012a,Sanna2011a,Sanna2013a,Prando2015,Surmach2017}.
On one hand, impurities act as pinning centers
and their presence is necessary for many devices. On the other
hand, impurities allow to probe the local response functions of
these materials and their effect on the superconducting transition
temperature T$_c$ and, accordingly,  to determine the symmetry of
the superconducting gap and the pairing mechanism. In the
iron-based superconductors (IBS) one of the most impressive
effects of impurities is associated with the substitution of Fe
with Mn in the LaFe$_{1-x}$Mn$_x$AsO$_{1-y}$F$_y$ family \cite{Sato2010a,Hammerath2014a}.
At optimal electron doping, namely for the $y$ values yielding the
maximum T$_c$, it was observed that a Mn concentration as low as
$x \gtrsim 0.001$ is enough to fully suppress superconductivity. The
disappearance of the superconducting phase takes place at a
quantum critical point where the spin correlation length diverges \cite{Hammerath2014a}
and for $x \gtrsim 0.001$ a stripe magnetic order arises \cite{Moroni2017}, the same type
of order characterizing LaFeAsO parent compound \cite{Maeter2009a,Lumsden2010}. Remarkably, at
the quantum phase transition a crossover in the resistivity
behavior is observed, evidencing an electronic localization \cite{Sato2010a,Kappenberger2018}. Such
a dramatic effect of a tiny amount of impurities indicates both
the presence of a very large local spin susceptibility, namely the
presence of strong electronic correlations, as well as a
collective effect of Mn impurities. In fact, a successful
explanation of the above phenomenology can be achieved by
considering the Ruderman-Kittel-Kasuya-Yosida (RKKY) coupling among Mn impurities which is
promoted by the enhanced spin susceptibility of the delocalized
electrons \cite{Gastiasoro2016,Moroni2017}.

The unique effect of Mn substitution was ascribed to the peculiar nature of the Mn$^{2+}$ ions, a $3d^5$ species characterized by a strong Hund's coupling \cite{Martinelli2017,Martinellisub}. Remarkably the first experimental evidence for the occurrence of a charge density wave state in IBS compounds was detected in the La(Fe,Mn)AsO system \cite{Martinelli2017}.

In the following we show that the changes in the
electronic structure driven by Mn impurities also cause a change in
the crystal structure and that, remarkably, Mn impurities tune the electronic, magnetic and structural properties
back to those observed in the parent compound LaFeAsO.

\section{Experimental and technical aspects}\label{sec:exp_details}
The \lafemnasof\ polycrystalline samples investigated here, with x = 0,
0.00025, 0.00075, 0.001, 0.002, 0.005, and 0.0075 are the same ones studied in Ref.~\cite{Hammerath2014a, Moroni2017}. We reported the characterization of the superconducting phase in Ref.~\cite{Hammerath2014a}, which are in agreement with previous data \cite{Sato2010a}.
We carried out high resolution X-ray diffraction measurements at the ID22 high-resolution powder diffraction beamline of the European Synchrotron Radiation Facility (ESRF) in Grenoble, France, at selected temperatures between 5 K and 300 K. These measurements are aimed at studying the presence of structural transition, suggested by previous nuclear quadrupole resonance measurements \cite{Moroni2017}.
We performed muon-spin spectroscopy ($\mu$SR) measurements on the GPS
instrument of the S$\mu$S facility at the Paul Scherrer Institute, Switzerland.
The muons act as nanoscopic magnetic sensors which allow to probe the spin dynamics and the local field arising
from the onset of a magnetic order \cite{Blundell1999a,Carretta2013}.
The full spin-polarization of the beams of positive muons
($\mu^{+}$) is the most peculiar feature exploited
by $\mu$SR. Accordingly, the great advantage of $\mu$SR with respect to other magnetic resonance techniques is that there is
no need to perturb the studied system with an external
polarizing magnetic field. In particular, we studied the local magnetism of \lafemnasof\
in conditions of external zero-magnetic field (ZF) (applied field $H_0$= 0).

\section{Results and Discussion}\label{sec:exp_details}

\subsection{Measurement of the structural transition}\label{ssec:structure}
\begin{figure}
\includegraphics[scale=0.09]{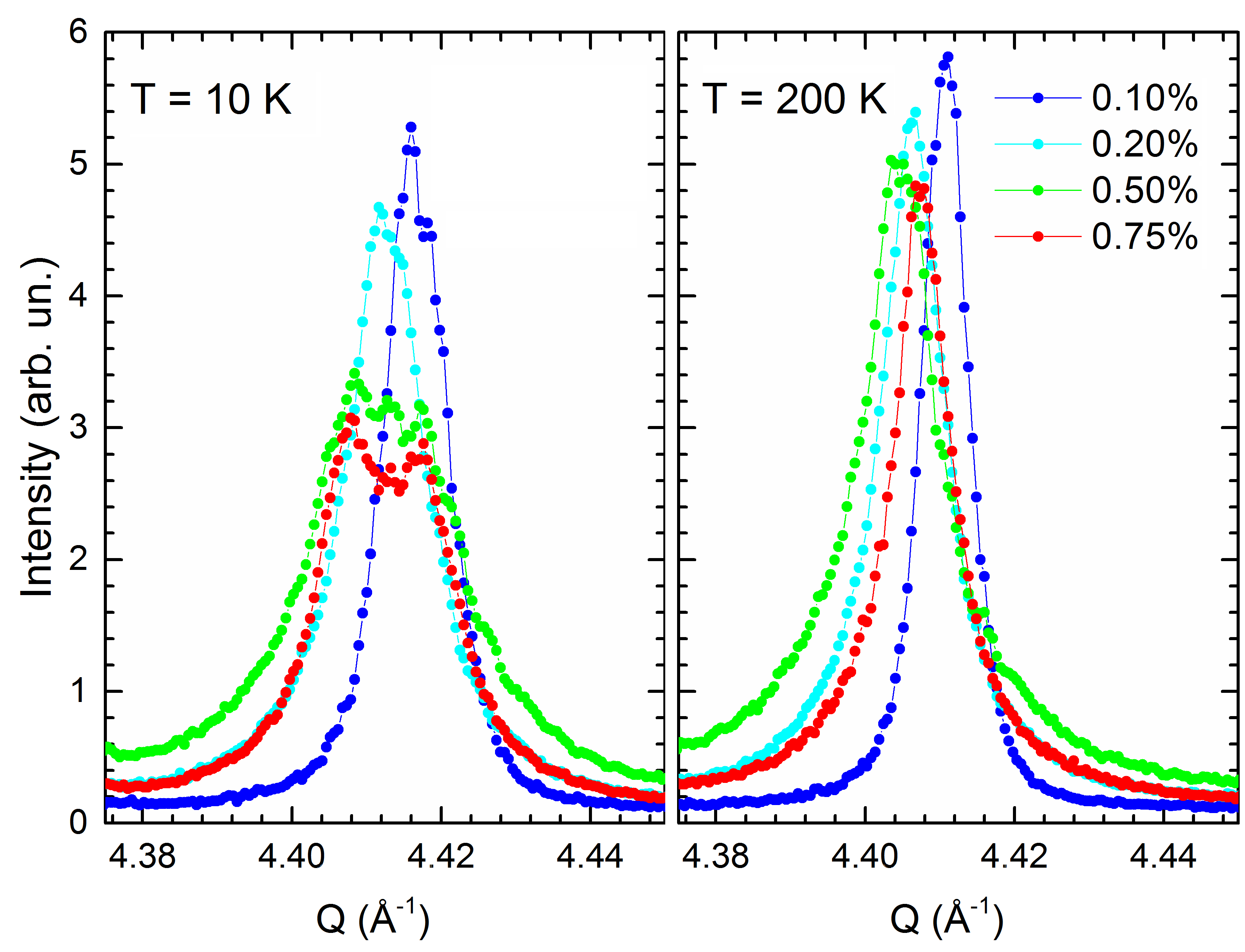} %xr-pattern
\caption{\label{fig:p220all} Superposition of diffraction patterns for \lafemnasof\ with $0.001\leq x\leq 0.0075$ in the Q-region where the tetragonal 220 peak is observed. On the right: data at 200 K; on the left: data at 10 K.}
\end{figure}
At room temperature all the samples display the standard \emph{P4/nmm} tetragonal structure as observed for the pure LaFeAsO compound. In particular, the compound LaFeAsO undergoes a tetragonal-to-orthorhombic structural transition on cooling at $\sim 150$ K \cite{Luetkens2009a,Frankovsky2013,Martinelli2016}. Nonetheless, this transition is progressively hindered by F-doping and hence the structural transition should be suppressed in LaFeAsO$_{0.89}$F$_{0.11}$). As a matter of fact, in our samples series the occurrence of the structural transition is dependent on the Mn-content, being progressively recovered with the increase of the Mn content. This is evidenced in Fig.~\ref{fig:p220all}, where the diffraction patterns collected for the inspected samples are superposed in the Q-region where the tetragonal 220 peak is observed (Q: momentum transfer, i.e. the modulus of the scattering vector).

At 200 K all samples display a single and quite symmetric peak, characterizing the tetragonal structure. At 10 K the peak split is evident for samples with $x = 0.005$ and 0.0075, indicating an orthorhombic crystal structure. For samples with $x = 0.002$ and 0.001 the peak is asymmetrically broadened; this feature can be possibly due an anisotropic distribution of the lattice microstrain in the \emph{hh0} plane or even to a convolution of orthorhombic diffraction lines producing an incomplete peak split (vide infra).

\begin{figure}[t!]
	\includegraphics[scale=0.15]{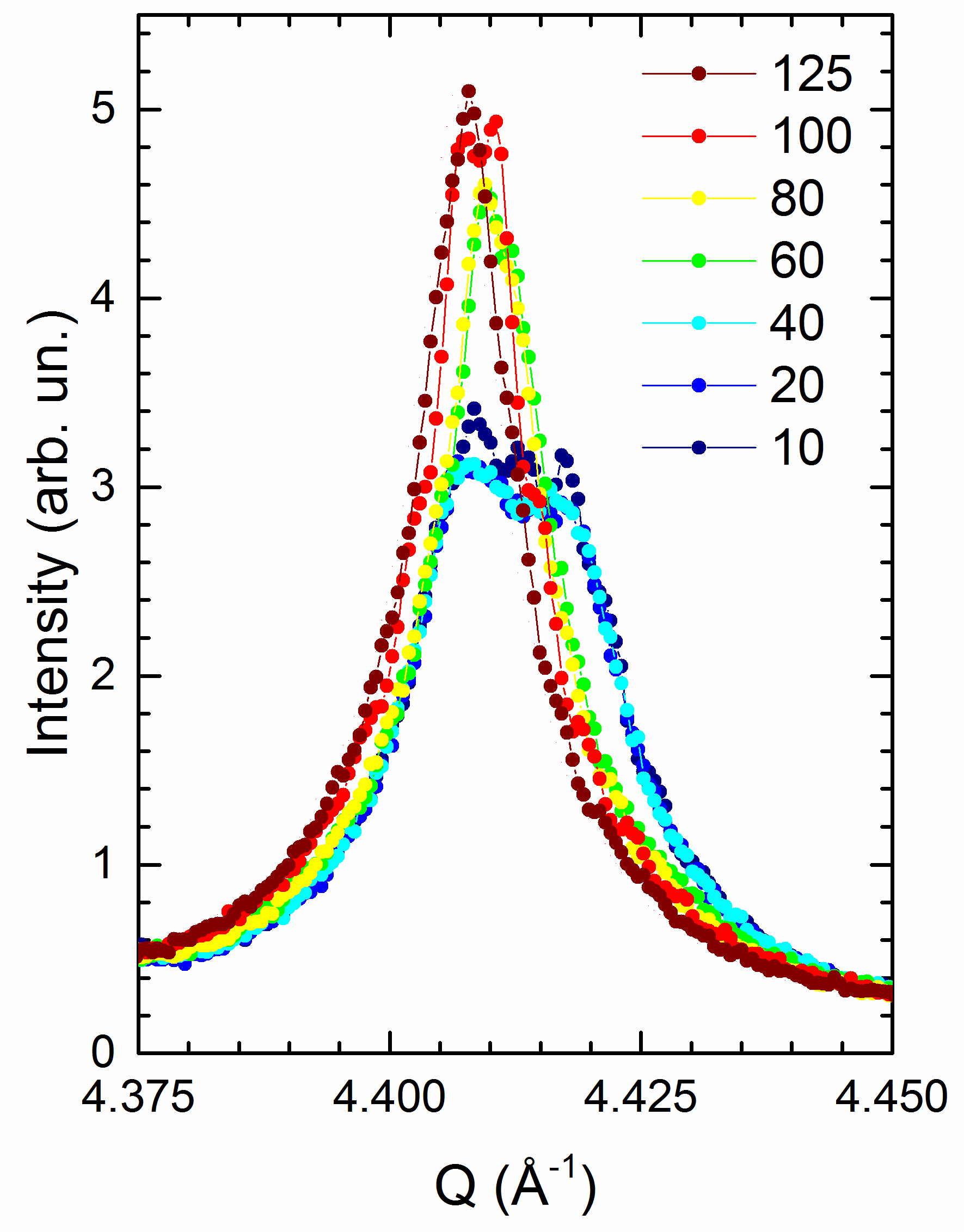} %xr-pattern
	\caption{\label{fig:p220vsT} Temperature evolution of the tetragonal \emph{220} peak, splitting into the orthorhombic \emph{400}+\emph{040} diffraction lines for \lafemnasof\ with $x=0.005$. The temperature list is in Kelvin.}
\end{figure}
The occurrence of the tetragonal-to-orthorhombic structural transition is generally based on the observation of selected peak splitting. Fig.~\ref{fig:p220vsT} shows the evolution on temperature of the tetragonal \emph{220} peak, splitting into the orthorhombic \emph{400}+\emph{040} diffraction lines at low temperature, as observed in the sample with $x = 0.005$; in this case, the peak split can be detected already at 100 K. As a matter of fact, a weak orthorhombic distortion prevents the formation of a resolved peak split, but rather produces a convolution of the orthorhombic diffraction lines into a single (although broadened) peak. In this case, careful structural and microstructural analysis are needed, in order to distinguish if broadening is originated by lattice microstrain or by a weak orthorhombic distortion. At this scope, the tetragonal and orthorhombic structural models were tested by Rietveld refinement and the structural transition temperatures $T_s$ can be confidently determined for the different samples (in this way we traced the tentative phase boundaries in the diagram of Fig.~\ref{fig:phasediagram} as a function of the Mn content). In particular, we analyzed the microstructural properties for both models, by refining the corresponding anisotropic strain parameters. If similar final  $\chi^2$ values resulted for both structural models (i.e. $\chi_{orthorhombic}$ slightly lower than  $\chi_{tetragonal}$), the significance test on the crystallographic \emph{R} factor was applied \cite{Hamilton1965} to determine the correct one.

\begin{figure}
\includegraphics[scale=0.13]{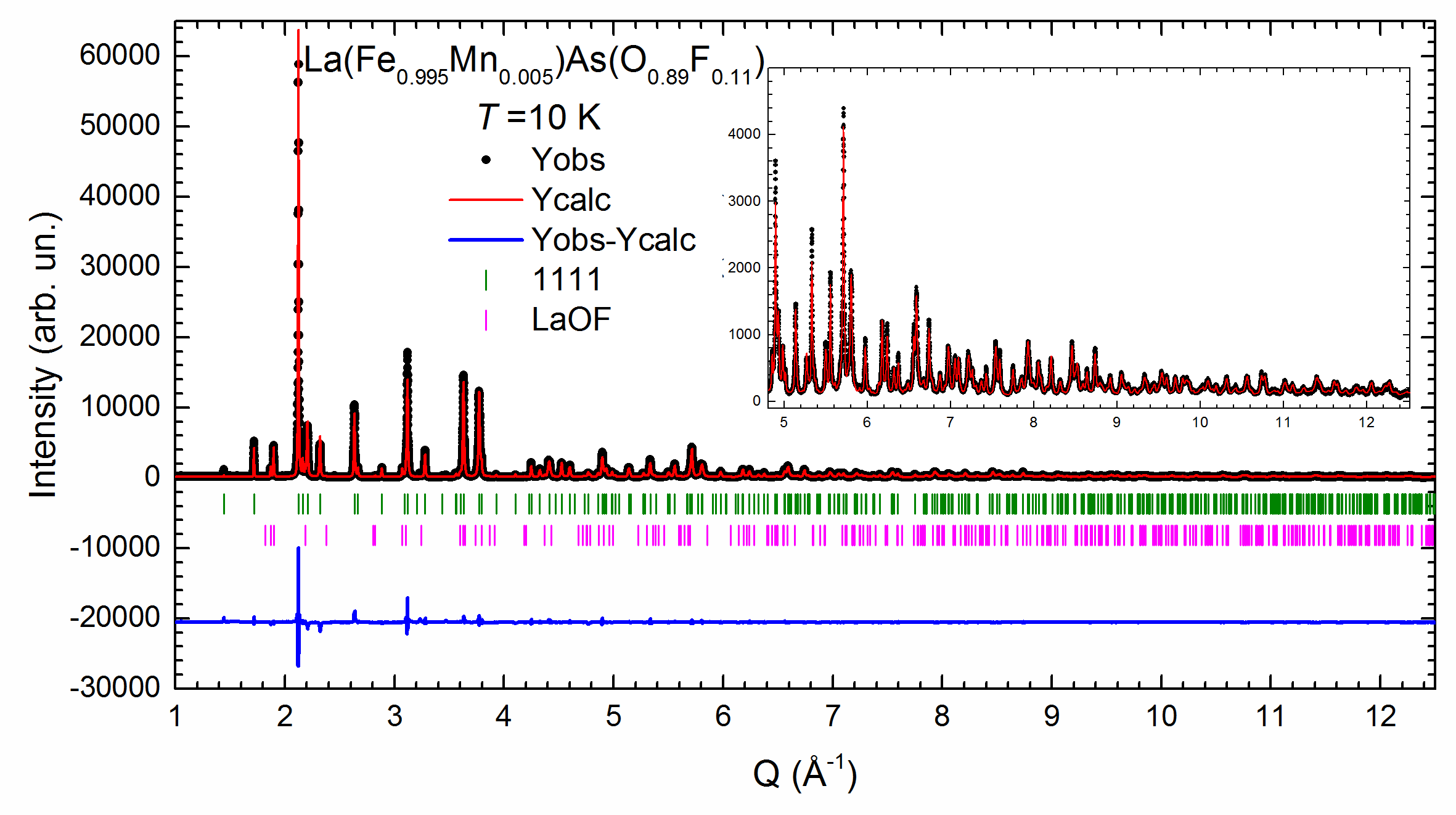}%xr-pattern
\caption{\label{fig:pattern} Rietveld refinement plot obtained for \lafemnasof\ with $x=0.005$ (data collected at 10 K); the inset shows an enlarged view of the high-Q data.}
\end{figure}

\begin{figure}[t!]
	\includegraphics[scale=0.7]{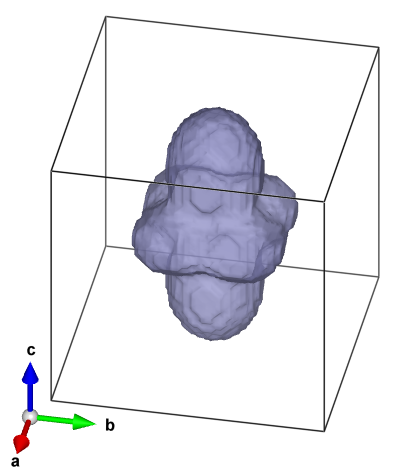}%xr-pattern
	\caption{\label{fig:isosurface} Observed tensor isosurface representing the microstrain broadening characterizing \lafemnasof\ with $x=0.001$ at 10 K.}
\end{figure}

Fig.~\ref{fig:pattern} shows the Rietveld refinement plot, selected as representative, obtained for data collected at 10 K for the sample with x = 0.005; the inset shows an enlarged view of the high-Q data.
The sample with $x = 0.001$ is located close to the boundary of the superconductive and the magnetic phase fields; for this reason, the exact characterization of its structural and microstructural properties is of particular interest. As a matter of fact, the reliability factors obtained after Rietveld refinement (data at 10 K) point to a tetragonal structure; nonetheless, the microstructural analysis reveals a significant in-plane microstrain. Fig.~\ref{fig:isosurface} shows the corresponding tensor isosurface, exhibiting an in-plane 4-fold tensor surface that is typically observed close to structural transitions involving a point group $4/mmm\rightarrow mmm$ transition \cite{Leineweber2011}. On these bases, two different scenarios can be proposed. In the first case, the structural transition takes place, but is not completed at 10 K and the sample is constituted by both tetragonal and orthorhombic polymorphs. In the second case, the underlying average structure is tetragonal, but short-range fluctuations of the lattice parameters produce a widespread and localized orthorhombic distortion.

\subsection{Measurement of the magnetic transition}\label{ssec:magnetism}

In a typical $\mu$SR experiment two opposite detectors count the number of positrons emitted along or in the opposite direction of the initial muon spin polarization, $N_F$ and $N_B$ respectively, arising from the asymmetric muon decay. The time evolution of the muon spin polarization is described by the muon asymmetry A(t) = ($N_F$ (t) - $N_B$(t)) / ($N_F$ (t) + $N_B$(t)) normalized by the total initial decay asymmetry $A_0$. The latter is calibrated for each sample in the paramagnetic phase.

\begin{figure}
\includegraphics[scale=0.15]{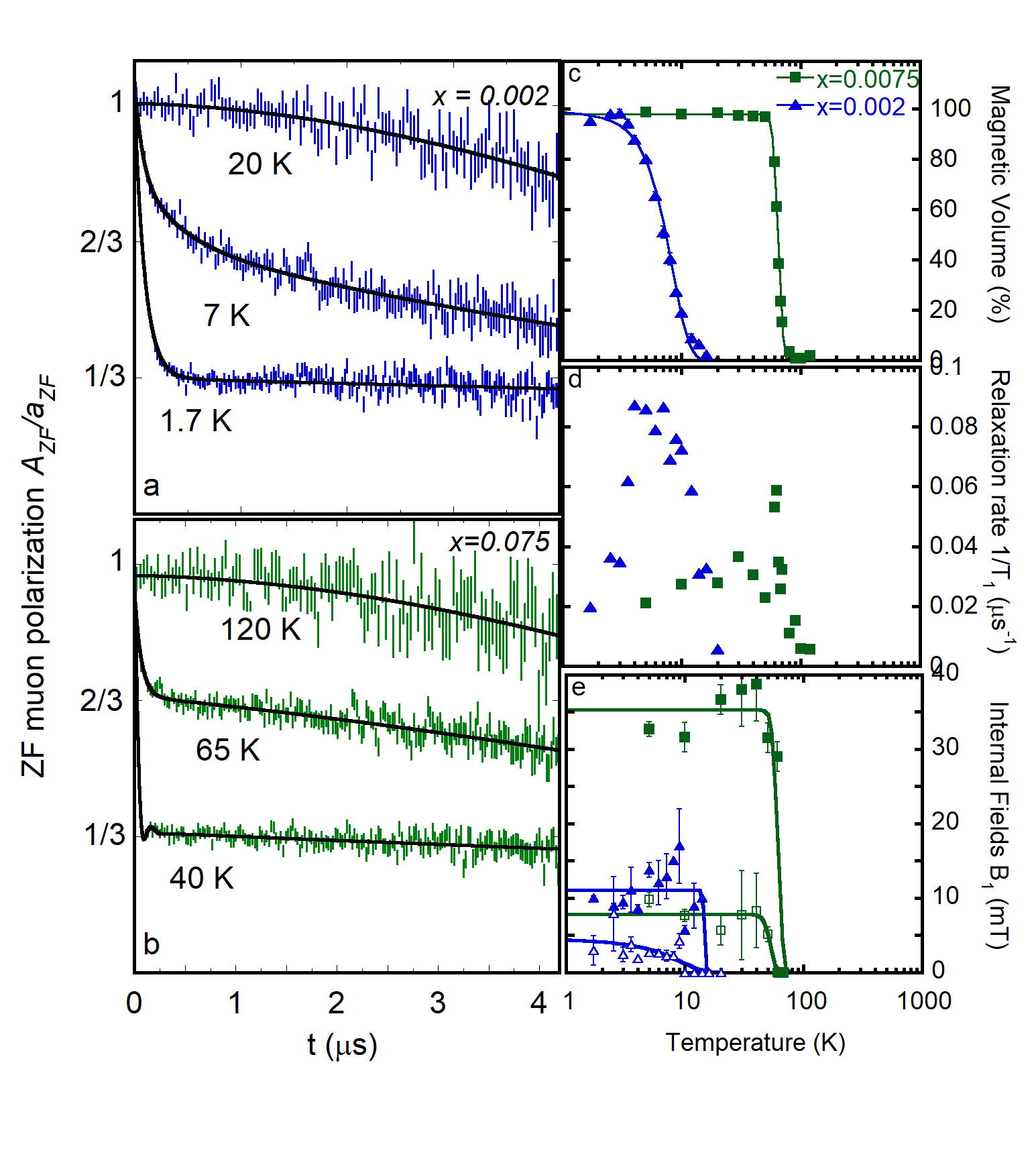} %ZFmuSR
\caption{\label{fig:zfmusr}representative ZF-muSR vs.\ temperature in  \lafemnasof\ for $x=0.002$ and $x=0.0075$.}
\end{figure}

No muon precessions are expected in a non magnetic phase (non magnetic samples or above the magnetic transition temperature) and the muon asymmetry A(t) is characterized by a single component with a slow gaussian decay rate ($\sim 0.1~\mu s^{-1}$), due to the interaction with the weak surrounding nuclear magnetic moments (Fig.\ref{fig:zfmusr}). Below the magnetic transition, each muon spin experiences a precession around a local field at t he muon site $B_\mu= \mathcal{A}_i \langle\mathbf{S}\rangle$, with $\mathcal{A}_i$ the hyperfine coupling tensor and $\langle\mathbf{S}\rangle$ the average Fe spin value, corresponding to the order parameter. The muon precession around $B_\mu$ gives rise either to an oscillating component of the muon asymmetry, when the precessions are coherent, or to a component with a fast decay rate in case of fast decoherence respect with to the period of the muon precession. The latter case is typical for a large local field distribution $\Delta B_\mu$, e.g. for short-range ordering or cluster spin ordering, characterized by overdamped oscillations of the muon signal for $\Delta B_\mu > B_\mu$.

For all the samples we fitted the muon polarization to the following phenomenological function:
\begin{eqnarray}
\label{eq:ZFAsymm}
      A(t)/A_0 & = & a_\perp \left[w_1 e^{-\lambda_1 t} f(\gamma_{\mu}B_{\mu,1}t)\right.\nonumber\\ & &
      \left. + w_2 e^{-\lambda_2 t} f(\gamma_{\mu}B_{\mu,2}t)\right]\nonumber\\ & & +
      a_{\parallel} e^{-t/T_1} e^{-(\sigma t)^2}
\end{eqnarray}
where $\gamma_{\mu}$ is the muon gyromagnetic ratio, $B_{\mu,i}$ for i=1 or 2 is the local field at two different muon sites, expected for 1111 compounds \cite{Carretta2013}, and $\lambda_{1,2}$ are the corresponding decay rates which reflect the muon field distribution $\Delta B_\mu$. The function $f(\gamma_{\mu}B_{\mu,i}t)$ is described by an oscillating function $f=cos(\gamma_{\mu}B_{\mu,i}t)$, in case of coherent muon precession, or by f=1 in case of overdamped oscillations of the muon signal for $\Delta B_\mu > B_\mu$. $a_{\parallel}$ and $a_{\perp}$, are the components of the muon spin parallel and perpendicular to the local field, which simple geometrical arguments for a fully magnetic powder sample predict to be 1/3 and 2/3, respectively. Hence we can calculate the magnetic volume as $V_{mag} = 3/2 a_\perp = 3/2 (1-a_\parallel)$.

A representative set of measurements of the time evolution of the muon asymmetry and the best fit to Eq.\ref{eq:ZFAsymm} for selected temperatures are displayed in Fig.\ref{fig:zfmusr}a and b for the samples with x=0.002 and 0.0075, respectively.
The temperature dependence of the magnetic volume fraction $V_{mag}(T)=
(3/2)(1- a_{\parallel}(T))$ is displayed in panel c which has been empirically fitted to $V_{mag}(T)=
0.5(1-erf(T- T_m^{av}/\sqrt{2}\Delta_V))$ (solid line), where $T_m^{av}$ represents an average magnetic transition temperature.
A peak of 1/$T_1$ due to critical fluctuations is observed when approaching the magnetic transition, as displayed in panel d.
Below the magnetic transition the size of spontaneous magnetic local fields at the muon sites $B_{\mu,i}$ can be directly determined from the fit of the oscillating component of the muon signal for x=0.0075. For the sample with x=0.002 we detected no oscillations and the size of the internal fields can be roughly determined as $B_\mu,i\sim \lambda_i/\gamma$. The temperature dependence of $B_\mu,i$ for i=1 and 2 is displayed in panel e for both the samples.
The temperature evolution of $V_{mag}$, $B_\mu$ and $1/T_1$ gives three independent ways to determine the magnetic transition temperature, which we are the same within about 2 K. The behavior of the magnetic transition as a function of the Mn doping is reported in Fig.~\ref{fig:phasediagram}.

\subsection{The phase diagram}\label{ssec:phasediagram}
\begin{figure}
\includegraphics[scale=0.17]{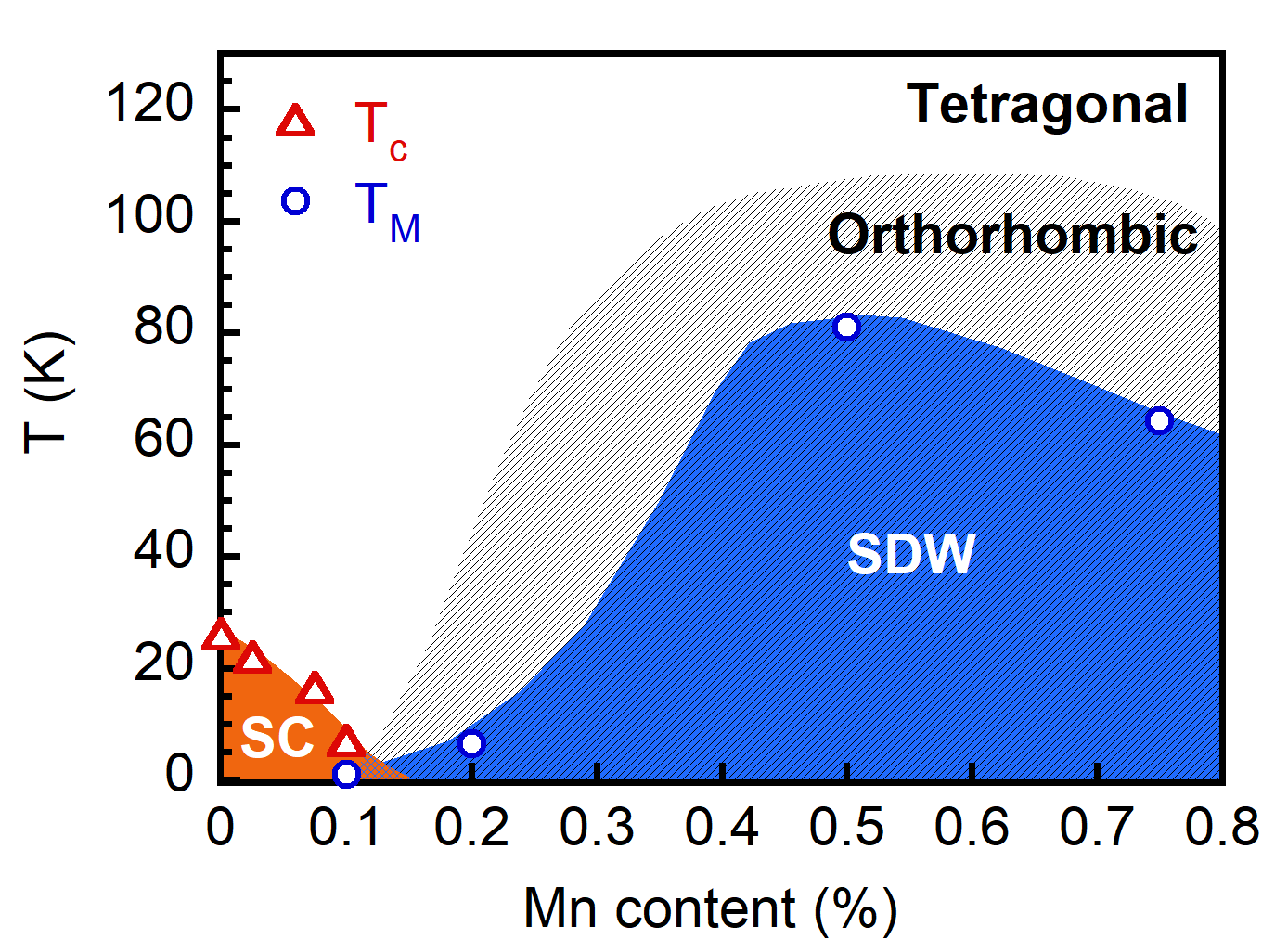} %ZFmuSR
\caption{\label{fig:phasediagram} Phase diagram as a function of Mn content for \lafemnasof\ displayng the superconducting, magnetic and structural phases.}
\end{figure}

We summarize the results of this work in the phase diagram displayed in Fig.~\ref{fig:phasediagram}.
The structural and microstructural characterization of the optimally electron doped \lafemnasof\ samples reveals that the tetragonal-to-orthorhombic structural transition is recovered for $x \gtrsim 0.001$. With the further enrichment with Mn, both the transition temperature and the amplitude of the orthorhombic distortion increase. Simultaneously the superconducting phase is suppressed and a static magnetic phase is induced. Recently a combination of nuclear magnetic resonance and \mossbauer spectroscopy \cite{Moroni2017} has shown that the magnetic structure of this phase is the ($\pi/a$, 0) stripe ordering typical of the spin density wave phase of the undoped LaFeAsO parent compound. This means that the substitution of Fe with an extraordinarily tiny amount of Mn ($\sim 0.1 \%$) has a dramatic effect on the electronic properties of the La1111 system. A similar qualitative behavior has been observed also in \lnfemnasof\ with \emph{Ln}= Sm \cite{Lamura2016} and La$_{0.8}$Y$_{0.2}$ \cite{Hammerath2015,Moroni2016, Kappenberger2018} but here a much higher amount of Mn, few percent, is needed to suppress superconductivity and induce static magnetism. In addition electric resistivity measurements \cite{Sato2010a,Kappenberger2018} and nuclear quadrupole resonance spectroscopy \cite{Moroni2016} have shown that a charge localization process is induced by Mn impurities in \emph{Ln}1111.

The effect of the Fe/Mn substitution in \emph{Ln}1111 has been theoretically studied by a five band real Hamiltonian model \cite{Gastiasoro2016,Moroni2017} which shows that the experimental results can be reproduced by considering two main ingredients: RKKY coupling among Mn impurities and electron correlation. The RKKY interaction is promoted by the enhanced spin susceptibility of the delocalized electrons and causes an enhanced spin polarization around Mn impurities. The model shows that the polarization mechanism is enhanced by increasing the electron correlation strength.

The comparison between experiments and theory \cite{Gastiasoro2016,Moroni2017} suggests that the electron correlation is reduced upon squeezing the lattice by increasing the chemical pressure with smaller radius \emph{Ln} ions, hence going from the bigger La to smaller Sm. We expect that the bandwidth of the system increases with the lattice squeezing, hence producing the enhancement of the kinetic energy and a weakening of the electron correlations. This mechanism qualitatively justifies why for \emph{Ln}=Sm the effect of Mn is weaker than for \emph{Ln}=La. These arguments suggest that the electron correlation strength in La1111 drives the system very close to an electronic instability. A small perturbation introduced by a tiny amount of Mn is enough to dramatically change the electronic properties of the system which at low temperatures precipitates back to the magnetic ($\pi/a$, 0) stripe spin density wave ground state of the parent compound with its associated orthorhombic structure.

\section{Conclusion}\label{sec:concl}
Based on muon spin relaxation measurements and high resolution X-ray diffraction, we draw the phase diagram of on \lafemnasof for $x=0-0.0075$ (Fig.~\ref{fig:phasediagram}). The superconducting phase is suppressed for a tiny amount of Mn doping $x\gtrsim0.001$ and both the magnetic ($\pi/a$, 0) stripe spin density wave phase and structural orthorhombic transition typical of the parent LaFeAsO compound are fully recovered.
The structural and microstructural characterization of the optimally electron doped \lafemnasof samples reveals that the tetragonal-to-orthorhombic structural transition is recovered for $x \gtrsim 0.001$. With the further addition of Mn, both the transition temperature and the amplitude of the orthorhombic distortion increase. These structural and microstructural properties underlie the crossover between the magnetic and the superconducting electronic ground states characterizing the \lafemnasof system. For the magnetic phase we considered a model \cite{Gastiasoro2016,Moroni2017} where RKKY interaction is promoted by the enhanced spin susceptibility of the delocalized electrons and polarizes the surrounding Fe spins. This model suggests that the fast change of the electronic and structural properties in La1111 is due to the presence of electronic correlations which drive the \emph{Ln}1111 system close to an electronic instability.

\section*{Acknowledgements}
This work was performed at the Swiss Muon Source S$\mu$S, Paul Scherrer Institut (PSI, Switzerland) and at The ESRF Facility (Grenoble, France). Authors acknowledge A. Fitch for his kind support during data collection at the ID22 beamline of ESRF (proposal HS-2646) and A. Amato and H. Luetkens for their kind support during the $\mu$SR experiment at PSI. Y. Kobayashi and M. Sato are kindly acknowledged for providing the sample compounds.

\end{document}